\documentclass[aps,pre,twocolumn,footinbib,showpacs,letterpaper,amsmath,amssymb,nobalancelastpage]{revtex4}
\usepackage{graphicx}
\usepackage{bm}
\usepackage{amsmath,amssymb}
\newcommand{ \ie }{\textit{i.e.} }
\newcommand{ \eg }{\textit{e.g.} }
\newcommand{ \e }[1]{\ensuremath{e^{\scriptstyle #1}} }
\newcommand{ \li }[2]{\ensuremath{ \mathrm{Li}_{#1}\left( #2 \right)}}
\newcommand{ \mean }[1]{\ensuremath{\left\langle #1 \right\rangle}}
\newcommand{ \orderof }[1]{\ensuremath{\mathcal{O}(#1)}}
\newcommand{ \cov }[2]{\ensuremath{\mathrm{cov}\{#1,#2\}}}
\newcommand{ \dtot }[3]{\ensuremath{ \frac{d^{#3} #1}{d #2^{#3}}}}
\newcommand{ \dpart }[3]{\ensuremath{ \frac{\partial^{#3} #1}{\partial #2^{#3}}}}

\newcommand{ \tr }[1]{\ensuremath{\text{Tr} \left( \bm{#1} \right)}}
\newcommand{\tk}{\ensuremath{\bm{\widetilde{k}}}}
\begin{document}
\title{Heterogeneous Bond Percolation on Multitype Networks with an Application to Epidemic Dynamics}
\author{Antoine Allard}
\affiliation{D\'epartement de physique, de g\'enie physique et d'optique, Universit\'e Laval, Qu\'ebec, Qu\'ebec, Canada G1V 0A6}
\author{Pierre-Andr\'e No\"el}
\affiliation{D\'epartement de physique, de g\'enie physique et d'optique, Universit\'e Laval, Qu\'ebec, Qu\'ebec, Canada G1V 0A6}
\author{Louis J. Dub\'e}
\affiliation{D\'epartement de physique, de g\'enie physique et d'optique, Universit\'e Laval, Qu\'ebec, Qu\'ebec, Canada G1V 0A6}
\author{Babak Pourbohloul}
\affiliation{University of British Columbia Centre for Disease Control, Vancouver, British Columbia, Canada V5Z 4R4}
\affiliation{School of Population \& Public Health, University of British Columbia, Vancouver, British Columbia, Canada V6T 1Z3}
\date{\today}
\begin{abstract}
Considerable attention has been paid, in recent years, to the use of networks in modeling complex real-world systems. Among the many dynamical processes involving networks, propagation processes --- in which final state can be obtained by studying the underlying network percolation properties --- have raised formidable interest. In this paper, we present a bond percolation model of multitype networks with an arbitrary joint degree distribution that allows heterogeneity in the edge occupation probability. As previously demonstrated, the multitype approach allows many non-trivial mixing patterns such as assortativity and clustering between nodes. We derive a number of useful statistical properties of multitype networks as well as a general phase transition criterion. We also demonstrate that a number of previous models based on probability generating functions are special cases of the proposed formalism. We further show that the multitype approach, by naturally allowing heterogeneity in the bond occupation probability, overcomes some of the correlation issues encountered by previous models. We illustrate this point in the context of contact network epidemiology.
\end{abstract}
\maketitle
%
% ***************************************************************************************************
\section{Introduction}
The end of the XXth century has witnessed increasing interest among the scientific community for the use of complex networks \cite{albert02_rmp,boccaletti06_pr,dorogovtsev01_aip,newman03_siamrev}
as models for many real-world systems, both from empirical and theoretical perspectives. From the empirical point of view, scientists have studied \emph{real-world} networks to highlight universal topological properties such as the Small-World effect \cite{watts98_nature}, highly skewed degree distributions \cite{albert99_nature,jeong00_nature,liljeros01_nature,redner98_epjb,vasquez02_pre} or assortative mixing \cite{newman02_prl}. On the theoretical side, models have been developed to describe or explain topological properties of networks \cite{barabasi99_science,newman01_pre,park04a_pre,pastor-satorras04}, to simulate their evolution in time \cite{gross08_jrsi} and the  dynamical processes taking place on them \cite{albert00_nature,moreno04b_pre,newman02_pre,pastor-satorras01_prl,young03}.

The first  models were rather simple: indistinguishable nodes joined by randomly placed edges \cite{erdos59_pmd}. With increasing information on real-world networks, more realistic --- and thus more complex --- models have been proposed taking into account properties such as an arbitrary degree distribution \cite{newman01_pre}, clustering \cite{newman03b_pre,serrano06_prl}, degree correlation \cite{newman03a_pre}, weighed edges \cite{newman04c_pre,yook01_prl}, directed edges \cite{bianconi08_prl,boguna05_pre,leicht08_prl,meyers06_jtb,newman01_pre,restrepo08_prl,schwartz02_pre} or mixing patterns \cite{catanzaro04_physA,newman02_prl}. Except for a few cases (\eg bipartite networks \cite{newman01_pre}), many existing models still consider only one type of nodes and therefore neglect any information characterizing the differences among the constituents of the simulated system. However, especially in social networks, these differences (\eg sex, age, ethnic group) may have significant and non-trivial effects on the structure (\eg assortative mixing, communities) and on the dynamical property(ies) of the networks themselves as well as on the dynamics of the phenomena of interest (such as disease propagation) throughout the networks \cite{gross06_prl,miller07_pre}.

In this paper, we present a bond percolation formalism of multitype networks with an arbitrary joint degree distribution where nodes have explicit properties associated with the type they belong to. On the one hand, the use of multitype networks allows one to reproduce mixing among nodes such as assortative mixing \cite{newman03a_pre} or clustering \cite{newman03b_pre}. On the other hand, the use of heterogeneous bond occupation probability allows one to take into account correlations between the probability of occupation of edges and the nature of the nodes they connect. When applied to epidemic propagation, we argue that this model adequately represents percolation (spreading) processes where such correlations are observed (\eg infectious diseases whose probability of transmission is correlated with intrinsic physiological and behavioral characteristics of individuals). 

Our paper is organized as follows. In Sec. \ref{sec:multi_net}, we introduce the multitype networks and define several quantities of interest. The formalism is developed in Sec. \ref{sec:pgf_formalism} where we obtain the occupied degree (and excess degree) distributions, the small component sizes, the percolation threshold, and the giant and (average) small component sizes. We also show that our formalism corresponds to a generalization of existing approaches \cite{meyers03_eid,newman02_pre,newman03a_pre} reducing to known results in the appropriate limits. This theoretical section is validated with a number of numerical simulations and followed by an application to epidemic dynamics in Sec. \ref{sec:appl_epid}, where the previously calculated quantities are interpreted in an epidemiological context. We also take the opportunity to explain how the proposed approach can overcome some of the correlation issues that should appear in a realistic 
treatment of epidemic propagation. Our conclusions and final remarks are then collected in the last section.
%
%
%
% ***************************************************************************************************
\section{Multitype Networks} \label{sec:multi_net}
\begin{figure}[tbp]
\includegraphics[width = 7.0cm]{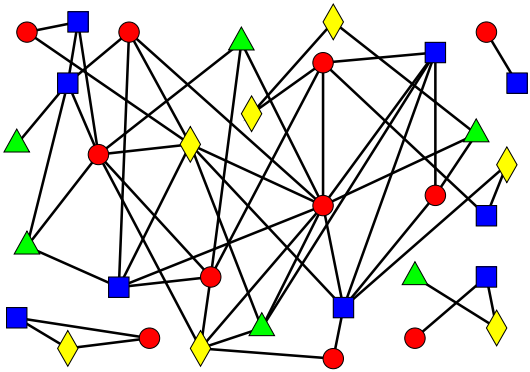}
\caption{\label{fig:multitype_networks} (Color online) Schematic representation of an undirected \emph{multitype} network with $M=4$, $N = 33$, $w_1 = \frac{3}{11}$, $w_2 = \frac{1}{3}$, $w_3 = \frac{2}{11}$ and $w_4 = \frac{7}{33}$ where types 1,2,3 and 4 refer to squares, circles, triangles and diamonds respectively. Edges running between nodes are bidirectional and  can thus be followed in either directions.}
\end{figure}

We consider undirected \emph{multitype} networks \cite{newman03a_pre,vasquez06_pre} defined as undirected networks composed of $N$ nodes, each of which are labeled with one of $M$ possible types. Type-$i$ nodes occupy a fraction $w_i$ of the network and the connections between nodes are prescribed by the degree distribution $P_i(k_1,k_2,\ldots,k_M) \equiv P_i(\bm{k})$ giving the joint probability for a randomly chosen type-$i$ node to be connected to $k_1$ type-1 nodes, $k_2$ type-2 nodes, $\ldots$ , $k_M$ type-$M$ nodes. Any mixing patterns between nodes such as assortative mixing are incorporated in the model via $P_i(\bm{k})$. Our networks are considered in the limit of large systems ($N \rightarrow \infty$) and are totally random in all respects other than the joint degree distribution $P_i(\bm{k})$ 
\footnote{We consider {\it simple} networks where no more than one edge can exist between two nodes and where there is no edge connecting a node to itself. The two possibilities have a probability of the order of \orderof{N^{-2}} and \orderof{N^{-1}} respectively in large random networks of size $N$.}. Therefore, $P_i(\bm{k})$ and $w_i$ define a network ensemble over which all quantities obtained with our formalism are averaged. Fig. \ref{fig:multitype_networks} shows an example of an undirected multitype network.

We now define $z_{ij}$ as the average number of edges leaving a type-$i$ node to type-$j$ nodes, directly obtained from $P_i(\bm{k})$ as
\begin{align} \label{eq:z_ij}
z_{ij} = \sum_{k_1=0}^{\infty} \ldots \sum_{k_M=0}^{\infty} k_j P_i(k_1,\ldots,k_M) \equiv \sum_{\bm{k=0}}^{\bm{\infty}} k_j P_i(\bm{k}).
\end{align}
Even if every edge in our networks is undirected, the presence of different types of nodes adds an \emph{artificial} direction to edges. Indeed, one can follow a link from a type-$i$ node to a type-$j$ node (noted $i \rightarrow j$) or in the opposite direction ($j \rightarrow i$). Since the degree distribution $P_i(\bm{k})$ prescribes the number of edges \emph{leaving} type-$i$ nodes, a given edge joining a type-$i$ and a type-$j$ node will be considered from two different perspectives. Therefore, to guarantee the consistency of the network ensemble, $P_i(\bm{k})$ and $w_i$ must respect the condition
\begin{align} \label{eq:wz_wzdagger}
\bm{w}\bm{z} = (\bm{w}\bm{z})^{T}
\end{align}
where
\begin{align*}
\bm{w} = \begin{bmatrix}   w_1  &    0   & \ldots &   0     \\
                            0   &   w_2  & \ldots &   0     \\
                         \vdots & \vdots & \ddots & \vdots  \\
                            0   &    0   & \ldots &  w_M    \end{bmatrix}; \quad
\bm{z} = \begin{bmatrix} z_{11} & z_{12} & \ldots & z_{1M}  \\
                         z_{21} & z_{22} & \ldots & z_{2M}  \\
                         \vdots   &   \vdots   & \ddots & \vdots      \\
                         z_{M1} & z_{M2} & \ldots & z_{MM}  \end{bmatrix}
\end{align*}
when $N \rightarrow \infty$. This constraint relies on having as many edges of type $i \rightarrow j$ as of type $j \rightarrow i$ \footnote{In practice, for realistic data, there will be as many edges of type $i \rightarrow j$ as of type $j \rightarrow i$ and \eqref{eq:wz_wzdagger} will naturally be respected.}. Note that \eqref{eq:wz_wzdagger} implies $\binom{M}{2}$ non-trivial generally overdetermined equations. Thus, one must use values for $P_i(\bm{k})$ and $w_i$ that explicitly fulfill \eqref{eq:wz_wzdagger}. The case $M=2$ is special in that $w_i$ can be uniquely determined from $P_i(\bm{k})$
\begin{align*}
w_1 = \frac{z_{21}}{z_{12} + z_{21}} \ ,\quad w_2 = \frac{z_{12}}{z_{12} + z_{21}}
\end{align*}
with $\tr{\bm{w}} = 1$.

Having now defined networks where one can identify the type of nodes, we are able to apply different probabilities of occupation according to how edges are followed. Thus, instead of having only one edge occupation probability $T$, as in other percolation models \cite{newman02_pre}, we define the bond occupation probability matrix
\begin{align} \label{eq:T}
\mathbf{T} = \begin{bmatrix} T_{11} & T_{12} & \ldots & T_{1M} \\
                             T_{21} & T_{22} & \ldots & T_{2M} \\
                             \vdots & \vdots & \ddots & \vdots \\
                             T_{M1} & T_{M2} & \ldots & T_{MM} \end{bmatrix}
\end{align}
where $T_{ij}$ is the occupation probability of the $i \rightarrow j$ edges. Note that $\mathbf{T}$ does not need to be symmetric and the probability of occupation of $i \rightarrow j$ edges can vary between edges of the same type (\ie linking the same ordered pair $ij$) as long as those values are independent identically distributed (\emph{iid}) random variables. The value of $T_{ij}$ is then simply the mean of their distribution \cite{newman02_pre} and is totally independent of $T_{ji}$, which is the mean of a different and independent distribution.

In view of the possible asymmetry of the probability of occupation, our approach is somewhat different from the traditional bond
percolation treatment which assumes a symmetric $\mathbf{T}$. 
It would perhaps be more appropriate to refer to our system as a {\it semi-directed bond percolation}.  
This denomination stems from the following point of view: one formally replaces every edge of the original undirected
network by two directed edges running in opposite directions and then uses the corresponding probability of occupation 
for each directed edges. This leads to a semi-directed network  whose percolation properties are easier to analyse. 
Therefore, the introduction of multitypes together with the tranmissibility matrix allows us to cover systems ranging
from classical bond percolation (symmetric $\mathbf{T}$) to spreading processes (asymmetric symmetric $\mathbf{T}$)
where directionality (e.g. causality) is implicitly present.
On this basis, we develop the multitype formalism in the next section.

%
% ***************************************************************************************************
\section{Formalism} \label{sec:pgf_formalism}
We now present a formalism that describes the heterogeneous bond percolation of multitype networks. It is based on probability generating functions (PGF) \cite{wilf94} and is a generalization to multitype networks of the formalism developed earlier by Newman \cite{newman02_pre}.
\subsection{Occupied Degree Distribution}
The first quantity needed to describe the percolation properties is the occupied degree distribution $\widetilde{P}_i(\tk)$, \ie the distribution of the number of occupied edges leaving a randomly chosen type-$i$ node. Assuming independence in the edges' occupation state, the probability that a randomly chosen \mbox{degree-$\bm{k}$} node has $\tk$ occupied edges is
\begin{align} \label{eq:P_i_given_k}
P_i(\tk|\bm{k}) = \prod_{l = 1}^{M} \binom{k_l}{\widetilde{k}_l} (T_{il})^{\widetilde{k}_l} (1-T_{il})^{k_l-\widetilde{k}_l}.
\end{align}
The probability that a randomly chosen node has $\tk$ occupied edges is then simply
\begin{align}
\widetilde{P}_i(\tk) & = \sum_{\bm{k} = \tk}^{\infty} P_i(\tk|\bm{k}) P_i(\bm{k}) \nonumber \\
            & = \sum_{\bm{k} = \tk}^{\infty} P_i(\bm{k}) \prod_{l = 1}^{M} \binom{k_l}{\widetilde{k}_l} (T_{il})^{\widetilde{k}_l} (1-T_{il})^{k_l-\widetilde{k}_l}, \label{eq:P_i_tilde}
\end{align}
where the summation convention is defined in (\ref{eq:z_ij}) here covering the ranges $\widetilde{k}_l \le k_l \le \infty$
for $1 \le l \le M$.
This probability is generated by the PGF $G_i(\bm{x};\mathbf{T}) \equiv G_i(x_1,\ldots,x_M;\mathbf{T})$
\begin{align}
G_i(\bm{x};\mathbf{T}) & = \sum_{\bm{\tk}=\bm{0}}^{\bm{\infty}} \widetilde{P}_i(\tk) \prod_{l=1}^{M}x_l^{\widetilde{k}_l} \nonumber \\
                       & = \sum_{\bm{k}=\bm{0}}^{\bm{\infty}} P_i(\bm{k}) \prod_{l = 1}^{M} \sum_{\widetilde{k}_l = 0}^{k_l} \binom{k_l}{\widetilde{k}_l} (x_lT_{il})^{\widetilde{k}_l} (1-T_{il})^{k_l-\widetilde{k}_l} \nonumber \\
                       & = \sum_{\bm{k}=\bm{0}}^{\bm{\infty}} P_i(\bm{k}) \prod_{l = 1}^{M} \big[ 1 + (x_l - 1)T_{il} \big]^{k_l}. \label{eq:G_i}
\end{align}
We see that $G_i(\bm{1};\mathbf{T})=1$ if $P_i(\bm{k})$ is properly normalized. We can obtain the average occupied degree $\widetilde{z}_{ij}$, \ie the average number of occupied edges leaving a type-$i$ node to \mbox{type-$j$} nodes, by using the differentiation property \cite{newman01_pre} of generating functions
\begin{align}
\widetilde{z}_{ij} & = \dtot{G_i(\bm{1};\mathbf{T})}{x_j}{} \nonumber \\
               & = T_{ij} \sum_{\bm{k}=\bm{0}}^{\bm{\infty}} k_j P_i(\bm{k}) \nonumber \\
               & = T_{ij} z_{ij} \label{eq:z_ij_tilde}
\end{align}
where $z_{ij}$ is the average degree defined by \eqref{eq:z_ij}.
\subsection{Occupied \emph{Excess} Degree Distribution}
Another useful and accessible quantity in our formalism is the occupied \emph{excess} degree distribution $\widetilde{Q}_{ij}(\tk)$. The \emph{excess} degree is defined as the number of edges leaving a node that have been reached by following a randomly chosen edge. For undirected \emph{unitype} networks ($M=1$), this quantity is simply the node's degree minus one (the edge that has already been followed). More information is required for multitype networks; one needs to know the type of node at both ends of the followed edge to correctly calculate the excess degree. This quantity is proportional to $k_iP_j(\bm{k})$ since high degree nodes are more likely to be reached from a randomly chosen edge than low degree nodes. Assuming independence in the occupation state of edges, the occupied excess degree distribution of a type-$j$ node reached from an $i \rightarrow j$ edge is given by
\begin{multline} \label{eq:Q_ij_tilde}
\widetilde{Q}_{ij}(\tk) = \frac{1}{z_{ji}} \sum_{\bm{k} = \tk}^{\bm{\infty}} \left(k_{i}+1\right) P_j(\bm{k} + \bm{\delta_i}) \\ \times \prod_{l=1}^{M} \binom{k_l}{\widetilde{k}_l} (T_{jl})^{\widetilde{k}_l} \left(1-T_{jl}\right)^{k_l-\widetilde{k}_l}
\end{multline}
where $P_j(\bm{k} + \bm{\delta_i}) \equiv P_j(k_1 + \delta_{i1},\ldots,k_M + \delta_{iM})$
and $\delta_{ij}$ is the delta of Kr\"onecker. Defining $F_{ij}(\bm{x};\mathbf{T})$ as the generating function associated with this distribution, we have
\begin{align*}
F_{ij}(\bm{x};\mathbf{T}) & = \sum_{\tk = \bm{0}}^{\bm{\infty}} \widetilde{Q}_{ij}(\tk) \prod_{l=1}^{M}x_l^{\widetilde{k}_l} \\
                          & = \frac{1}{z_{ji}} \sum_{\bm{k} = \bm{0}}^{\bm{\infty}} k_{i} P_j(\bm{k}) \prod_{l=1}^{M} \big[ 1 + (x_l - 1)T_{jl} \big]^{k_l-\delta_{il}}
\end{align*}
which can also be obtained from $G_{i}(\bm{x};\mathbf{T})$ by differentiation
\begin{align} \label{eq:F_ij}
F_{ij}(\bm{x};\mathbf{T}) = \frac{1}{\widetilde{z}_{ji}} \dtot{G_{j}(\bm{x};\mathbf{T})}{x_i}{}
\end{align}
where $\widetilde{z}_{ij}$ is the average occupied degree defined by \eqref{eq:z_ij_tilde}.
\subsection{Small Components Size Distribution} \label{sec:small_comp_size_dist}
\begin{figure}[tb]
\begin{center}
\includegraphics[width = 0.9\linewidth]{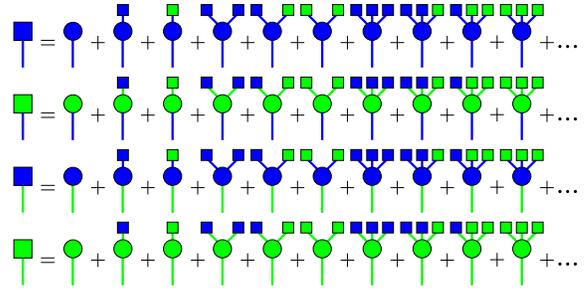}
\caption{\label{fig:contributions} (Color online) Illustration of the consistency relation \eqref{eq:H_ij} for $M=2$. The boxes represent the component reached by an $i \rightarrow j$ edge and the circles stand for the type-$j$ node first reached. The color of the edges refer to the type of the node from which one has arrived and the color of the box/circle stands for the type of the node reached first.}
\end{center}
\end{figure}

We now wish to calculate the size distribution of \emph{small} components in the network ensemble. A component is any closed set (cluster) of nodes connected by occupied edges. The adjective small is meant to qualify any \emph{intensive} component (\ie one that does not scale with the network size). Let us first define $H_{ij}(\bm{x};\mathbf{T})$ as the function generating the size distribution of the component reached by following an $i \rightarrow j$ edge. Small components are typically finite, except at the phase transition where their average size diverges \cite{stauffer94}. Thus, we expect the probability of finding closed loops in finite components to go as \orderof{N^{-1}}, which is negligible in the large system size limit ($N \rightarrow \infty$). 
Small components are therefore treelike in structure and $H_{ij}(\bm{x};\mathbf{T})$ can be decomposed in an additive set of contributions as graphically shown at Fig. \ref{fig:contributions} for the case $M=2$. 
The size distribution of a small cluster reached from an $i \rightarrow j$ edge arises from two situations:
either the edge reaches a node that has no outgoing occupied edges (i.e. occupied excess degree = 0), 
or it reaches a node that has outgoing occupied edges (i.e. occupied excess degre $\not=$ 0) that lead to other clusters whose size distribution is given by $H(\bm{x};\mathbf{T})$ as well
\footnote{In an \emph{infinite} network, $H_{ij}(\bm{x};\mathbf{T})$ is invariant under translation on the network, \ie one always sees the same small components size distribution independently where he/she stands in the small component.}.

Noting that the distribution of outgoing edges is given by $F_{ij}(\bm{x};\mathbf{T})$ and using the \emph{power} property \cite{newman01_pre} of generating functions leads to the consistency relation
\begin{align} \label{eq:H_ij}
H_{ij}(\bm{x};\mathbf{T}) = x_j F_{ij}\big(\bm{H_{j}}(\bm{x};\mathbf{T});\mathbf{T}\big)
\end{align}
where the right-hand side of the equation must be read as $F_{ij}\big(H_{j1}(\bm{x};\mathbf{T}),\ldots,H_{jM}(\bm{x};\mathbf{T});\mathbf{T}\big)$. The solution to \eqref{eq:H_ij} is found by seeking the stable fixed point of the mapping
\begin{align*}
H_{ij}^{(n)}(\bm{x};\mathbf{T}) = x_j F_{ij}\big(\bm{H^{(n-1)}_{j}}(\bm{x};\mathbf{T});\mathbf{T}\big)
\end{align*}
as $n \to \infty$ for initial conditions $H^{(0)}_{ij}(\bm{x};\mathbf{T}) = x_j$. Technically, the existence of a stable fixed point is guaranteed by the presence of the
$x_j$ factor in \eqref{eq:H_ij}. Indeed, it implies that the coefficients in front of variables whose powers sum to $n$ \mbox{(\eg $x_1^{k_1}\ldots x_M^{k_M} \text{ with } \sum_{l=1}^{M}k_l = n$)} are  exact after precisely $n+1$ iterations.

Let us next consider a randomly chosen type-$i$ node. Each of its leaving edges leads to a component whose size distribution is generated by $H_{ij}(\bm{x};\mathbf{T})$. Defining $K_i(\bm{x};\mathbf{T})$ as the function generating the size distribution of the whole component, we have
\begin{align} \label{eq:K_i}
K_i(\bm{x};\mathbf{T}) = x_i G_{i}\big(\bm{H_{i}}(\bm{x};\mathbf{T});\mathbf{T}\big).
\end{align}
Since type-$i$ nodes occupy a fraction $w_i$ of the network, the size distribution of the component reached from a randomly chosen node is generated by
\begin{align}
K(\bm{x};\mathbf{T}) & = \sum_{i=1}^{M} w_i K_i(\bm{x};\mathbf{T}) \nonumber \\
                     & = \sum_{i=1}^{M} w_i x_i G_{i}\big(\bm{H_{i}}(\bm{x};\mathbf{T});\mathbf{T}\big). \label{eq:K}
\end{align}
Similar equations to \eqref{eq:H_ij} and \eqref{eq:K_i} have already been derived in \cite{newman03a_pre} to obtain the size distribution of the small component reached from a randomly chosen type-$i$ node. However, the PGFs used there were functions of only one variable instead of $M$ (\ie $x$ instead of $\bm{x}$) and therefore did not generate the composition of the small component (\ie the number of nodes of each type). Thus, \eqref{eq:H_ij} and \eqref{eq:K_i} are generalized versions of previoulsy derived expressions.
\subsection{Percolation Threshold}
Percolation is usually characterized by the divergence of the correlation length. This translates here in the divergence of the average size $\mean{s}$ of small components. Using the \emph{moments} property \cite{newman01_pre} of PGFs, the average number of type-$i$ nodes in the small component reached from a randomly chosen node is obtained by differentiating \eqref{eq:K} with respect to $x_i$
\begin{align} \label{eq:mean_s_i}
\mean{s_i} = w_i + \sum_{l=1}^{M} w_l \sum_{j=1}^{M} \widetilde{z}_{lj} \alpha_{lj}^{(i)}
\end{align}
where $\alpha_{lj}^{(i)} \equiv \left. \dtot{H_{lj}(\bm{x};\mathbf{T})}{x_i}{} \right|_{\bm{x}=\bm{1}}$ are the solutions of
\begin{align} \label{eq:alpha_lji}
\alpha_{lj}^{(i)} = \delta_{ij} + \sum_{n=1}^{M} T_{jn} \beta_{lj}^{(n)} \alpha_{jn}^{(i)}.
\end{align}
We have isolated in \eqref{eq:alpha_lji} the average number of type-$n$ nodes that can be reached from a type-$j$ node arrived at by following a $l \rightarrow j$ edge (average excess degree)
\begin{align} \label{eq:beta_ljn}
\beta_{lj}^{(n)} = \frac{1}{T_{jn}} \left. \dpart{F_{lj}(\bm{x};\mathbf{T})}{x_n}{}\right|_{\bm{x} = \bm{1}}
\end{align}
which only depends on the network structure (\ie its degree distribution) and is therefore known. Thus, to obtain $\mean{s_i}$, one simply has to solve \eqref{eq:alpha_lji}, $M$ sets of $M^2$ equations and $M^2$ unknowns. It can be shown that all $\alpha_{lj}^{(i)}$ are inversely proportional to $\det{(\mathbf{I}-\mathbf{A})}$ where $\mathbf{I}$ is the identity matrix and $\mathbf{A}$ is a $M \text{x} M$ block matrix whose blocks ($\mathbf{A_{ij}}$) are themselves $M \text{x} M$ matrices with
\begin{align} \label{eq:A_ij_munu}
[\mathbf{A_{ij}}]_{\mu\nu} = T_{ij}\beta_{\mu\nu}^{(j)} \delta_{i\nu}
\end{align}
giving the $(\mu,\nu)$ element of the $(i,j)$ block. For example, in the case $M=2$, $\mathbf{A}$ takes the form
\newlength{\bw}
\setlength{\bw}{7.5ex}
\begin{align}
\mathbf{A} = \begin{bmatrix} \\[-3.5ex]
\begin{bmatrix} \makebox[\bw]{$T_{11}\beta_{11}^{(1)}$} & \makebox[\bw]{$0$} \\ \makebox[\bw]{$T_{11}\beta_{21}^{(1)}$} & \makebox[\bw]{$0$} \end{bmatrix} &
\begin{bmatrix} \makebox[\bw]{$T_{12}\beta_{11}^{(2)}$} & \makebox[\bw]{$0$} \\ \makebox[\bw]{$T_{12}\beta_{21}^{(2)}$} & \makebox[\bw]{$0$} \end{bmatrix} \\[3 ex]
\begin{bmatrix} \makebox[\bw]{$0$} & \makebox[\bw]{$T_{21}\beta_{12}^{(1)}$} \\ \makebox[\bw]{$0$} & \makebox[\bw]{$T_{21}\beta_{22}^{(1)}$} \end{bmatrix} &
\begin{bmatrix} \makebox[\bw]{$0$} & \makebox[\bw]{$T_{22}\beta_{12}^{(2)}$} \\ \makebox[\bw]{$0$} & \makebox[\bw]{$T_{22}\beta_{22}^{(2)}$} \end{bmatrix} \\[3ex]
\end{bmatrix}.
\end{align}
From \eqref{eq:mean_s_i}, we see that the average size $\mean{s}$ of the component reached from a randomly chosen node  diverges as
\begin{align} \label{eq:mean_s}
\mean{s} = \sum_{i=1}^{M} \mean{s_i} \propto \frac{1}{\det{(\mathbf{I}-\mathbf{A})}}
\end{align}
for $\det{(\mathbf{I}-\mathbf{A})} \rightarrow 0$. Therefore the phase transition happens when $\det{(\mathbf{I}-\mathbf{A})} = 0$ which marks the point where the giant component first appears. This result is in accord with the corresponding expression found in \cite{newman03a_pre}
and, as noted earlier, is again more general.
\subsection{Giant Component}
Beyond the percolation threshold, there is an \emph{extensive} cluster (the giant component) in the network. In \mbox{Sec. \ref{sec:small_comp_size_dist}}, $H_{ij}(\bm{x};\mathbf{T})$ has been defined as generating the size distribution of \emph{finite} components. Thus, $H_{ij}(\bm{x};\mathbf{T})$, $K_i(\bm{x};\mathbf{T})$ and $K(\bm{x};\mathbf{T})$ are no longer normalized beyond the percolation threshold since it is not guaranteed that a randomly chosen edge/node will lead to a finite component (although a fraction of them may lead to the giant component). Therefore, the probability that a randomly chosen type-$i$ node leads to the giant component is simply
\begin{align}
\mathcal{P}_i & = 1 - K_i(\bm{1};\mathbf{T}) \nonumber \\
              & = 1 - G_i \big( \bm{\overrightarrow{h_{i}}};\mathbf{T} \big) \label{eq:P_i}
\end{align}
with $G_i \big( \bm{\overrightarrow{h_{i}}};\mathbf{T} \big) \equiv G_i \big( \overrightarrow{h_{i1}},\ldots,\overrightarrow{h_{iM}};\mathbf{T} \big)$. We have noted $\overrightarrow{h_{ij}} \equiv H_{ij}(\bm{1};\mathbf{T})$ (read $h_{ij}$ \emph{forward}) as the probability that a randomly chosen $i \rightarrow j$ edge leads to a finite component, and is the solution of
\begin{align} \label{eq:h_ij_direct}
\overrightarrow{h_{ij}} & = F_{ij}\big(\bm{\overrightarrow{h_{j}}};\mathbf{T}\big)
\end{align}
obtained by evaluating \eqref{eq:H_ij} at $\bm{x} = \bm{1}$. If one randomly chooses a node in the network, the probability that it leads to the giant component is therefore
\begin{align} \label{eq:P}
\mathcal{P} = \sum_{i=1}^{M} w_i \mathcal{P}_i = 1 - \sum_{i=1}^{M} w_i G_i \big( \bm{\overrightarrow{h_{i}}};\mathbf{T} \big).
\end{align}

To calculate the size of the giant component, one needs to know the probability that a randomly chosen node is {\it not} linked to the giant component by any of its edges (\ie that this node can not be reached from the giant component). One simple way to obtain this information is to study the network topology by following every edges \emph{backwards}. This can be achieved with our formalism by simply using $\mathbf{T}^T$ (the transpose of $\mathbf{T}$) instead of $\mathbf{T}$ since any given type-$i$ node can be left (reached) by any of its edges with the probability $T_{ij}$ ($T_{ji}$). We define $\overleftarrow{h_{ij}}$ (read $h_{ij}$ \emph{backward}) as the probability that a given type-$i$ node can not be reached from the giant component by a $j \rightarrow i$ edge. This quantity is calculated as solution of
\begin{align} \label{eq:h_ij_reverse}
\overleftarrow{h_{ij}} & = F_{ij}\big(\bm{\overleftarrow{h_{j}}};\mathbf{T}^T\big).
\end{align}
Therefore, we see that $G_i \big( \bm{\overleftarrow{h_{i}}};\mathbf{T}^T \big)$ is the probability that a randomly chosen type-$i$ node does not belong to the giant component. The fraction of the network occupied by type-$i$ nodes that are in the giant component is thus given by
\begin{align} \label{eq:S_i}
\mathcal{S}_i & = w_i\big[ 1 - G_i \big( \bm{\overleftarrow{h_{i}}};\mathbf{T}^T \big) \big]
\end{align}
and the size of the giant component is 
\begin{align} \label{eq:S}
\mathcal{S} = \sum_{i=1}^{M} \mathcal{S}_i = 1 - \sum_{i=1}^{M} w_i G_i \big( \bm{\overleftarrow{h_{i}}};\mathbf{T}^T \big).
\end{align}
In comparing \eqref{eq:P} and \eqref{eq:S}, one will see that asymmetry of the bond occupation probability matrix implies that $\mathcal{P} \neq \mathcal{S}$. This quantitative difference between $\mathcal{P}$ and $\mathcal{S}$ resides in the asymmetry in the number of occupied edges of type $i \rightarrow j$ and of type $j \rightarrow i$. A naive generalization of the formalism introduced in \cite{newman02_pre} would have missed the distinction between $\mathcal{P}$ and $\mathcal{S}$. Clearly for symmetric transmissibility $\mathbf{T} = \mathbf{T}^T$, one would have $\mathcal{P}=\mathcal{S}$. A similar result has previously been discussed in \cite{meyers06_jtb} for semi-directed networks and, with different approaches,
it has been obtained for {\it undirected} networks in \cite{miller07_pre,kenah07_pre}. The present demonstration
is a new extension to  the latter class of networks.
\subsection{Average Small Components Size}
Above the percolation threshold, $K(\bm{x};\mathbf{T})$ still generates the size distribution of the finite component reached from a randomly chosen node, although it needs to be normalized according to 
\begin{align*}
\frac{K(\bm{x};\mathbf{T})}{1-\mathcal{P}}
\end{align*}
since $\mathcal{P}\neq 0$. The general expression for $\mean{s_i}$ is therefore
\begin{align} \label{eq:mean_s_i_over}
\mean{s_i} = \frac{w_i G_{i}\big(\bm{\overrightarrow{h_{i}}};\mathbf{T}\big)}{1-\mathcal{P}} + \frac{1}{1-\mathcal{P}} \sum_{l=1}^{M} w_l \sum_{j=1}^{M} \widetilde{z}_{lj} \overrightarrow{h_{jl}} \alpha_{lj}^{(i)}
\end{align}
where $\alpha_{lj}^{(i)}$ is the average number of type-$i$ nodes in the finite component reached by a randomly chosen $l \rightarrow j$ edge and is the solution of
\begin{align} \label{eq:alpha_lji_over}
\alpha_{lj}^{(i)} = F_{lj}\big(\bm{\overrightarrow{h_{j}}};\mathbf{T}\big) \delta_{ij} + \sum_{n=1}^{M} T_{jn} \beta_{lj}^{(n)} \alpha_{jn}^{(i)}.
\end{align}
Analogously to \eqref{eq:beta_ljn},
\begin{align} \label{eq:beta_ljn_over}
\beta_{lj}^{(n)} = \frac{1}{T_{jn}} \left. \dpart{F_{lj}(\bm{x};\mathbf{T})}{x_n}{}\right|_{\bm{x}=\bm{\overrightarrow{h_{j}}}}.
\end{align}
One can see that \eqref{eq:mean_s_i_over}--\eqref{eq:beta_ljn_over} reduce to \eqref{eq:mean_s_i}--\eqref{eq:beta_ljn} in absence of the giant component since in this case $\mathcal{P} = 0$ and $\overrightarrow{h_{ij}} = 1$. Technically, \eqref{eq:mean_s_i_over}--\eqref{eq:beta_ljn_over} can be very useful to obtain information on the small components without having to solve \eqref{eq:H_ij}--\eqref{eq:K}, a very time consuming operation for large $M$ or for networks with large small components. It is also possible to calculate the second moments of $K(\bm{x};\mathbf{T})$
\begin{align} \label{eq:mean_s_i_s_j_over}
\mean{s_is_j} = \left.\frac{1}{1-\mathcal{P}}\dtot{}{x_j}{}\left[ x_i \dtot{K(\bm{x};\mathbf{T})}{x_i}{} \right] \right|_{\bm{x}=\bm{1}}
\end{align}
from which the covariance matrix of the small components size \mbox{$\big(\cov{s_i}{s_j} = \mean{s_is_j} - \mean{s_i}\mean{s_j}\big)$} is obtained. Clearly, iterative equations for $\mean{s_is_j}$,
 similar to \eqref{eq:mean_s_i_over} and \eqref{eq:alpha_lji_over}, can be derived to calculate the covariance matrix without solving \eqref{eq:H_ij}--\eqref{eq:K}. Higher moments can also be obtained in a similar way.
\subsection{Special Cases} \label{sec:special_cases}
We now show that, in corresponding limit cases, our formalism reproduces the already published theoretical results. Firstly, one can easily verify that all of the equations in the previous section reduce to the ones in \cite{newman02_pre} when $M=1$. Secondly, equations associated with the components (small or giant) in \cite{newman03a_pre} can be obtained by setting $T_{ij}=1 \ \forall \ i,j $ in our equations and $\bm{x}=x$ in \eqref{eq:H_ij} and \eqref{eq:K_i}. Thirdly, results obtained from a semi-directed formalism such as the one in \cite{meyers06_jtb} can also be obtained with our formalism by setting $T_{ij}=0$ for some $ij$ pairs while keeping $T_{ji} \neq 0$.
Fourthly, for bipartite networks ($M=2$), all edges are connecting different types of nodes and the
constraint 
\begin{align*}
P_i(k_1,k_2) = 0 \ \forall \ k_i \neq 0
\end{align*}
must be imposed, implying that
\begin{equation*}
\begin{split}
z_{ii} = 0;\quad \beta_{ii}^{(j)} = 0;\quad \beta_{ji}^{(i)} = 0;\quad \alpha_{ii}^{(j)} = 0.
\end{split}
\end{equation*}
 $F_{ii}(x_1,x_2;\mathbf{T})$ and $H_{ii}(x_1,x_2;\mathbf{T})$ are then undefined. From \eqref{eq:A_ij_munu}, we see that the phase transition in this case, $\det{(\mathbf{I}-\mathbf{A})}=0$, occurs when
\begin{align*}
T_{12}T_{21}\beta_{12}^{(1)}\beta_{21}^{(2)} = 1 \ ,
\end{align*}
a result previously obtained in \cite{meyers03_eid,newman02_pre}. Moreover, one can obtain the average number of type-1 nodes in the component reached from a randomly chosen type-1 node $\mean{s_1}_1$ under the percolation threshold by differentiating \eqref{eq:K_i} with respect to $x_i$ and solving \eqref{eq:alpha_lji_over}
\begin{align*}
\mean{s_1}_1 = 1 + \frac{T_{12} T_{21} z_{12} \beta_{12}^{(1)}}{1 - T_{12}T_{21}\beta_{12}^{(1)}\beta_{21}^{(2)}},
\end{align*}
a result also obtained by the authors of \cite{meyers03_eid,newman02_pre}. Furthermore, it is possible to calculate the size of the giant component as in \cite{meyers03_eid} by setting $x_2=1$ in \eqref{eq:H_ij} and \eqref{eq:K_i} with the constraints listed above. 

An even more general constraint $P_i(\bm{k}) = 0 \ \forall \ k_i \neq 0$ can be used to obtain a formalism for multipartite networks. Our approach can therefore incorporate clustering effects by assigning some of the node types to groups and then using the projected network (where nodes belonging to the same group are linked together) as proposed in \cite{newman03b_pre}.
%
%
%
% ***************************************************************************************************
\section{Application to Epidemiology} \label{sec:appl_epid}
Over the years, mathematical models \cite{andersonmay91,hethcote00_siamrev} have provided insights on the factors influencing diseases propagation dynamics and have improved testing intervention/prevention strategies. Outbreaks of respiratory pathogens (\eg SARS \cite{meyers05_jtb}) and sexually transmitted infections (STIs) have encouraged the emergence of models using network theory to capture the patterns of potential disease-causing interactions between individuals \cite{newman02_pre,meyers06_jtb,moreno02_epjb,pastor-satorras01_prl}. Despite the many successes, most of these models are still based on a simplifying assumption, which limits the realistic simulation of disease propagation for certain categories of diseases.
Before discussing how the quantities obtained from our formalism can be translated in an epidemiological setting, we briefly
 state  some of the difficulties associated with a realistic epidemic dynamics and the possible advantage of a multitype description.
\subsection{Failure of the \emph{iid} hypothesis and heterogeneity} \label{sec:iid_hyp}

The \emph{iid} hypothesis \cite{newman02_pre} assumes that the probability of transmission between any pair of individuals is an \emph{independent identically distributed} random variable taken from a given distribution. Thus, the \emph{a priori} probability of transmission, $T$, between any two individuals is the mean of this distribution  and, in the population as a whole, the disease will propagate from an infectious individual to a susceptible one with the same probability $T$. This implies that no correlations, whatsoever, can be taken into account.

However, the probability of transmission of infectious diseases is typically dependent on intrinsic immunological and behavioral traits of individuals. Many infectious diseases show heterogeneity in their transmissibility. For example, the human immunodeficiency virus (HIV) has a higher efficiency of transmission from male to female than female to male \cite{nicolosi94_epid,glynn01_aids}. There is also strong evidences that co-infection with other STIs could facilitate HIV transmission \cite{vernazza99_aids,grosskurth95_aids,wasserheit92_std}. In regards to influenza, it has been shown that 
children (under 15 years old) are more likely to transmit the disease than adults \cite{cauchemez04_statmed}. Further, it has been shown that the \emph{iid} hypothesis fails to adequately model  \emph{susceptible-infectious-recovered} (SIR) dynamics when the distribution of the infectious period $P(\tau)$ is not sharp around a given value $\tau_0$ \cite{kenah07_pre,noel08_arxiv}. Therefore, most existing percolation approaches fail to realistically simulate the propagation of some infectious diseases due to the inability of the \emph{iid} hypothesis to model the correlations between individual's traits (including their infectious period) and the probability of transmission.

If one could identify specific individuals within the network, one could determine \emph{who infects whom} and 
it would become possible to apply the appropriate probability of transmission. Hence, difficulties raised by the heterogeneity 
in  transmissibility could be largely overcome by considering node heterogeneity. This suggests that,
in order to properly model the propagation of a large class of diseases, one could separate the nodes into a sufficient number of categories (types) 
to insure that the \emph{iid} hypothesis can  be applied correctly. Our multitype formalism can then be used to investigate the percolation properties of the corresponding system. 

Confronted with a situation where the infectious period is broadly distributed and heterogeneity is present, one could
adopt the following line of action.  In the case of influenza, one could split the population between adults and children;  or between male and female when modeling HIV propagation. The probability of transmission could still vary according to the \emph{iid} hypothesis, within the same type of edges, if nodes are separated into a sufficient number of groups, within each of which all significant correlations are explicitly included. The finite width of the infectious period distribution $P(\tau)$ can be accounted for by  
simply dividing its contribution into a sufficient number of duration subdomains $[\tau_{i-1},\tau_{i}]$ 
(each associated with a node type randomly distributed in the population if more detailed information is not available) and using the corresponding transmissibility in our model. The fraction of the network occupied by type-$i$ nodes will then be $w_i = \int_{\tau_{i-1}}^{\tau_i} P(\tau) d\tau$. The same procedure is also applicable if the susceptibility of individuals is heterogeneous.
\subsection{Epidemiological quantities}
We now interpret the quantities that can be calculated with our formalism in an epidemiological context. The contact network topology is prescribed by $P_i(\bm{k})$ and $w_i$ while the bond occupation probability matrix entries, $T_{ij}$, are the average probability of transmission from infectious individuals of type $i$ to susceptible individuals of type $j$. $K_i(\bm{x};\mathbf{T})$ generates the outbreak size distribution caused by patient zero (\ie the first known individual to 
become infected who directly or indirectly causes all subsequent infections) of type-$i$ (\eg adult, child; male, female). Similarly, $K(\bm{x};\mathbf{T})$ generates the outbreak size distribution caused by a patient zero of any type. The quantity $\mean{s_i}$ is the average number of type-$i$ individuals infected from patient zero and $\mean{s}=\sum_{i=1}^{M}\mean{s_i}$ is the expected size of an outbreak. One could also differentiate \eqref{eq:K_i} with respect to $x_j$ to obtain the average number of type-$j$ individuals infected by a patient zero of type-$i$ (see Sec. \ref{sec:special_cases}). Those quantities can be useful, for example, in evaluating the impact of strategies focused on the reduction of  morbidity in specific population groups (\eg health care workers, the elderly).

For a given contact network, $\det{(\mathbf{I}-\mathbf{A})}$ is a polynomial in powers of the elements of $\mathbf{T}$ whose coefficients depend only of the network topology. Thus, the percolation threshold, $\det{(\mathbf{I}-\mathbf{A})}=0$, defines the \emph{critical transmissibility set} over which there is a non-zero probability that an outbreak will turn into a large-scale epidemic. The probability that such an epidemic will occur is given by $\mathcal{P}$ and by $\mathcal{P}_i$ if patient zero is known to be of type-$i$. Should an epidemic occur, the fraction of the population that will eventually be infected 
is given by $\mathcal{S}$ while $\mathcal{S}_i$ indicates the fraction of the population of infected individuals of type $i$. Note that if an outbreak dies out while having infected only a finite number of individuals (or a small number compared to the size of the population), the expected number of infected individuals of type $i$ is still given by $\mean{s_i}$ computed
  with \eqref{eq:mean_s_i_over} (this remark holds for $\mean{s}$ as well).
\subsection{Numerical Simulations}
To illustrate how our formalism could be applied in an epidemiological context and to confirm its predictions, we have performed extensive computer simulations on multitype networks of $N=10^{5}$ nodes divided into two types \mbox{($M=2$)}. We have considered a contact network where the distribution of the total degree ($k_1+k_2$) of individuals is given by a power-law with an exponential cutoff and where the probability that an edge leaving a type-$i$ node arrives on a type-$j$ node is given by $p_{ij}$. Thus, the joint degree distribution of our network is
\begin{align*}
P_i(k_1,k_2) = \frac{(k_1+k_2)^{-\eta_{i}} \e{-(k_1+k_2)/\kappa_{i}}}{\li{\eta_{i}}{\e{-1/\kappa_{i}}}} \cdot \binom{k_1+k_2}{k_1} p_{i1}^{k_1} p_{i2}^{k_2}
\end{align*}
with the parameters
\begin{align*}
\bm{\eta} = \left[ \begin{array}{c} 1 \\ 2 \end{array} \right]; \qquad \bm{\kappa} = \left[ \begin{array}{c} 8 \\ 10 \end{array} \right] ; \qquad \bm{p} = \left[ \begin{array}{cc} 0.7 & 0.3 \\ 0.4 & 0.6 \end{array} \right].
\end{align*}
$\li{\eta}{z}$ denotes the $\eta$th polylogarithm of $z$ \cite{lewin81} (also known as Jonqui\`ere's function). We 
have used a  simple joint degree distribution to
\begin{figure}[tb]
\begin{center}
\includegraphics[width = 0.95\linewidth]{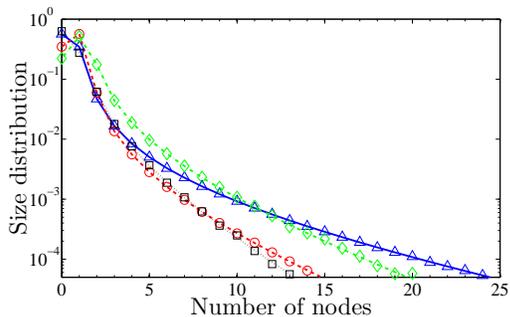}
\caption{\label{fig:size_distribution}(Color online) Size distribution of small components obtained by numerical simulations (symbols) compared with the theoretical prediction of \eqref{eq:K} (lines).  $\triangle$ and  $\lozenge$ correspond to the number of type-1 nodes in small components $\big(\text{generated by }K(x_1,1;\mathbf{T})\big)$ for $\gamma=0.1$ and $\gamma=0.5$ respectively.  $\bigcirc$ and  $\square$ are the equivalent quantities but for type-2 nodes $\big(\text{generated by }K(1,x_2;\mathbf{T})\big)$.}
\end{center}
\end{figure}
illustrate our point; nonetheless our formalism is very general and $P_i({\bm{k}})$ could include many non-trivial correlations as shown in \cite{newman03a_pre}. To show the effect of the asymmetry of $\mathbf{T}$ on $\mathcal{P}$ and $\mathcal{S}$, we have used the following transmissibility matrix
\begin{align*}
\mathbf{T} = \gamma \left[ \begin{array}{cc} 0.95 & 0.98 \\ 0.48 & 1.00 \end{array} \right]
\end{align*}
where $\gamma$ allows us to vary the infectiousness of the disease. 
The specific choice of the elements of $\mathbf{T}$ has no particular relevance here, except perhaps to result
in large  $\mathcal{P}$ and $\mathcal{S}$ values for $\gamma = 1$. By solving $\det{(\mathbf{I}-\mathbf{A})}=0$ for $\gamma$, we find that the epidemic transition occurs when $\gamma_c \simeq 0.1834$.

\begin{figure}[tbp]
\begin{center}
\includegraphics[width = 0.95\linewidth]{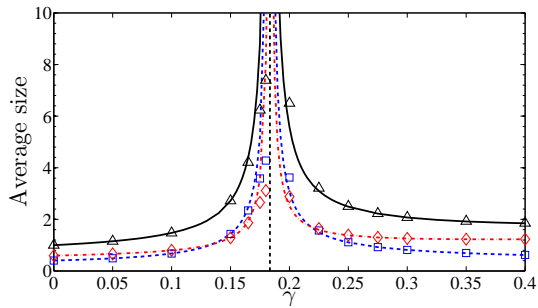}
\caption{\label{fig:average_size}(Color online) Average number of nodes in small components as predicted by \eqref{eq:mean_s_i_over} (lines) compared to simulations results (symbols; $\square$: type-1 nodes,  $\lozenge$: type-2 nodes and  $\triangle$: both) as a function of $\gamma$. The vertical dashed black line indicates the percolation threshold ($\gamma_c$).}
\end{center}
\end{figure}

We have generated 2000 multitype networks following a method similar to the one described in \cite{newman03a_pre}, with the degree distribution presented above. We have then performed epidemic simulations by infecting a randomly chosen node and 
allowing the disease to propagate with probabilities given by $\mathbf{T}$. 
Above the percolation threshold, we have identified the components (small or giant) by setting 
a size-parameter, a percentage of the total number of nodes, below which the cluster was registered to belong
to the set of small components. Experimentation has shown the final results rather insensitive to the exact value
of the size-parameter and we have settled conservatively for a value of 0.5\% of $N$. 
Figure \ref{fig:size_distribution} compares the distribution of the number of infected nodes of each type caused by an outbreak predicted by \eqref{eq:K}, with the results of numerical simulations under ($\gamma=0.1$) and above ($\gamma=0.5$) the epidemic threshold. 
One observes a very good agreement between the theoretical prediction and the simulations. This quantitative accord
(in this figure and the following ones)
is representative of a much larger set of calculations carried out with different values of the transmissibility matrix elements. 
Figure \ref{fig:average_size} shows the average number of infected nodes in an outbreak for each type of node and for different values of $\gamma$. Theoretical predictions are obtained from \eqref{eq:mean_s_i_over}. Again, an excellent agreement between our model predictions and numerical simulations is recorded; 
the small disagreement around the percolation threshold is caused by the finite size of the networks used for the simulations. 
Indeed as $N$ decreases, finite size effects  become important and the formalism would have to be modified along the lines 
described in \cite{noel08_arxiv}, for instance. Preliminary results  indicate however that agreement between
results of the present formalism and numerical simulations is maintained, even if the size of the network is reduced 
to $N=1000$. A more extensive study of the issue of finite size in a multitype network is under investigation.

\begin{figure}[tbhp]
\begin{center}
\includegraphics[width = 0.95\linewidth]{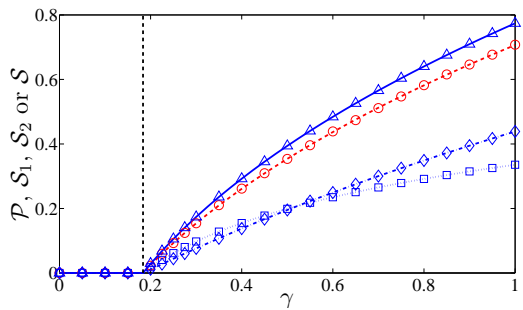}
\caption{\label{fig:giant_comp}(Color online) Probability of reaching the giant component from a randomly chosen node ($\mathcal{P}$,  $\bigcirc$), fraction of the network occupied by type-1 nodes ($\mathcal{S}_1$,  $\square$), type-2 nodes ($\mathcal{S}_2$,  $\lozenge$) and both node types ($\mathcal{S}$,  $\triangle$) in the giant component. Lines stand for theoretical predictions and symbols for simulation results. The vertical dashed black line indicates the percolation threshold ($\gamma_c$).}
\end{center}
\end{figure}

Finally, Fig. \ref{fig:giant_comp} compares the values predicted by our model for the probability of an epidemic to occur ($\mathcal{P}$) and its relative size ($\mathcal{S}_1$, $\mathcal{S}_2$ and $\mathcal{S}$) with simulation results for different values of $\gamma$. Again, there is a very good agreement between the theoretical predictions and results from simulations. The asymmetry of $\mathbf{T}$ is responsible for the significant difference between $\mathcal{P}$ and $\mathcal{S}$ (up to approximately 10\% in this case). We also see that the presence of node types allows more detailed information on the final state of an epidemic since it is then possible to determine the number of individuals of each 
type that are infected during an epidemic. Moreover, Fig. \ref{fig:giant_comp} demonstrates that these numbers do not 
remain proportional for varying transmissibilities. To the best of our knowledge, such information was not possible to obtain in previous percolation models. Furthermore, the heterogeneity in transmissibility in our formalism allows one to test more specific public-health policies.
For instance, one could study the effectiveness of age-specific influenza control strategies such as vaccination, face masks or hand-washing by varying the transmissibility matrix entries for the relevant age groups.
Therefore, the multitype approach of our model offers more detailed information on outbreak outcomes. This is very useful when comparing the cost-effectiveness of prevention or intervention strategies.
%
%
%
% ***************************************************************************************************
\section{Conclusion}
In this paper, we have introduced a bond percolation formalism on multitype networks. The formalism explicitly allows heterogeneity in the edge occupation probability via the matrix $\mathbf{T}$ whose elements $T_{ij}$ are the probability for an $i \rightarrow j$ edge to be occupied. Using probability generating functions (PGF), we have
obtained several exact forms of classical statistical properties (in the limit of large networks) such as, the size distribution of small components, the probability of reaching the giant component from any node as well as its relative size. Furthermore, the presence of node types has allowed us to obtain more detailed information on the composition of small components, the giant component, and a general expression for the percolation threshold. We have also obtained iterative equations for the average number of type-$i$ nodes in the small component, which allows to easily and rapidly obtain information on the network structure.

We have also shown that our model is a generalized version of various existing approaches based on the PGF. Many known results and effects can be obtained with our model. For instance, equations describing the bond percolation of multipartite networks can easily be derived from our formalism. While semi-directed networks have been previously used to simulate the asymmetry between population groups infecting each other \cite{meyers06_jtb}, this effect can be achieved with our 
undirected network model by setting $T_{ij}=0$ for some $ij$ pairs while keeping $T_{ji} \neq 0$. Thus, type-$j$ nodes will be able to infect type-$i$ nodes, while  
transmission in the other direction will not be possible. A completely general semi-directed extension of our formalism (with $3M$ variables, say $\bm{x}$, $\bm{y}$, $\bm{z}$, for the 3 ways to move across the network, 
following the links forward, backward and in both directions)
is straightforward to derive. This extended formalism would be required when the underlying network includes directed edges whose presence  can not be randomly
determined with a probability $T_{ij}$ that solely depends on the edge type (here $i \rightarrow j$), \ie, 
additional correlations exist.

These structural properties have considerable influence on the dynamical processes taking place on networks; this, in turn, can have  a significant impact on their topology. Therefore, a formalism such as the one presented in this paper can be used to probe and characterize the structure (by setting $T_{ij}=1 \ \forall \ i,j = 1,\ldots,M$) of an evolving network at a given time in order to predict the network's topological evolution.

The approach described in this paper, when compared to previous methods, facilitates more realistic simulations of the propagation of infectious diseases manifesting heterogeneity in their transmissibility. 
We argue that heterogeneity in nodes is a way to overcome some correlation issues caused by heterogeneous transmissibility.
In addition, the presence of different types of nodes allows the simulation of many non-trivial mixing patterns 
observed in real-world networks, such as assortativity, the preferential connection between different types of nodes; and clustering, the fact that nodes belonging to a specific group are more likely to be connected to one another in the contact network.
 Thus, the proposed model is suitable for more detailed and more precise epidemiological investigations (\eg impact of intervention or prevention strategies on specific population groups) resulting in  more adapted and effective recommendations to public health authorities. Hopefully, models such as the one presented in this paper joined with ever increasing theoretical developments will contribute to the improvement of public health policies.
%
%
%
% ***************************************************************************************************
\section*{Acknowledgments}
BP acknowledges the support of the Canadian Institutes of Health Research (CIHR, grants no. MOP-81273 and PPR-79231), the Michael Smith Foundation for Health Research (Senior Scholar Funds) and the British Columbia Ministry of Health (Pandemic Preparedness Modeling Project). AA and PAN were partly supported by the above grants and PAN is also thankful to CIHR for further support. LJD is grateful to NSERC (Canada) and FQRNT (Qu\'ebec) for continuing support.
%
%

% \bibliography{networks}

%
%

\end{document}